\documentclass[aps,pra,floatfix,showpacs,twocolumn,superscriptaddress]{revtex4}
\usepackage{graphicx}
\usepackage{amsmath}
\usepackage{amssymb}
\usepackage{latexsym}
\usepackage{epsfig}
\usepackage{bbm}

 

\begin{document}
\title{Spontaneously axisymmetry breaking phase\\ in a binary mixture of spinor Bose-Einstein condensates}
\author{Z. F. Xu}
\affiliation{Department of Physics, Tsinghua University, Beijing
100084, People's Republic of China}
\author{J. W. Mei}
\affiliation{Institute for Advanced Study, Tsinghua University, Beijing
100084, People's Republic of China}
\author{R. L\"u}
\affiliation{Department of Physics, Tsinghua University, Beijing
100084, People's Republic of China}
\author{L. You}
\affiliation{Department of Physics, Tsinghua University, Beijing
100084, People's Republic of China}

\date{\today}

\begin{abstract}
We study the ground state phases for a mixture of two atomic spin-1 Bose-Einstein condensates (BECs)
in the presence of a weak magnetic (B-) field. The ground state is found to contain a
broken-axisymmetry (BA) phase due to competitions among intra- and inter-species
spin exchange interactions and the linear Zeeman shifts. This is in contrast to
the case of a single species spin-1 condensate,
where the axisymmetry breaking results from competitions among the
linear and quadratic Zeeman shifts and the intra-species ferromagnetic interaction.
All other remaining ground state phases for the mixture are found to preserve
axisymmetry. We further elaborate on the ground state phase diagram and
calculate their Bogoliubov excitation spectra. For the BA phase, there exist
three Goldstone modes attempting to restore the broken U(1) and SO(2) symmetries.
\end{abstract}


\maketitle

\section{Introduction}

The spin degrees of freedom is described by the SU(2) symmetry
group, which can take different irreducible representations,
each corresponding to different values of the spin.
For an atom in a hyperfine spin $f$ state, its corresponding SU(2)
representation is $2f+1$ dimension. With the help of optical traps,
spinor condensates with atoms in all these representations
have been realized in many different atomic species and states
\cite{ueda10,kurn98,stenger98,barret01,schmaljohann04,griesmaier05}.

Extensive experimental and theoretical investigations have
focused mainly on atomic spinor condensates with spin equal to 1,
2, and 3. For the simplest case of spin-1 condensates \cite{ho98,ohmi98,law98,reijnders04,mueller04},
the mean-field ground states \cite{ho98,koashi00}, and exact full quantum eigenstates \cite{ho00,koashi00}
are intensively explored, also noted is the possibility of
topological fractionalized 1/2-vortex excitations in the
anti-ferromagnetic phase \cite{zhou01} and the appearance and formation of spin domains
in the ferromagnetic phase \cite{zhang05,sadler06}.
In addition, theoretical studies \cite{law98,zhang05a,chang07}
either based on mean-field approximation or including quantum spin fluctuations,
and experimental efforts \cite{chang05,liu09} have been made to investigate
the spin dynamics of the condensate. Similar efforts are directed at
spin-2 condensates \cite{ciobanu00,koashi00,ueda02},
whose ground states contain one extra phase: the cyclic phase, with an
exotic symmetry described by the non-Abelian tetrahedral group T,
potentially of significant value for demonstrating various
topological excitations \cite{makela03,semenoff07,kobayashi09}.
Under the single spatial mode approximation
(SMA), the exact eigenspectra and eigenstates of a spin-2 condensate
are derived \cite{koashi00,ueda02}, which shows that the ground state
is in general fragmented.
The spin dynamics are also studied in the presence of a B-field, with the linear and quadratic Zeeman shifts
\cite{schmaljohann04,widera06,kronjager06}.
For the case of the spin-2 hyperfine manifold of a $^{87}$Rb condensate \cite{kronjager06},
its spin dynamics show amplitude resonance when the B-field is tuned experimentally
just like spin mixing in a spin-1 condensate \cite{zhang05a,chang07}.
When the spin increases to 3 \cite{griesmaier05}, the ground state contains
many more phases \cite{diener06,santos06}, and new possibilities for
richer and more colorful physics become available, waiting to be further explored.

Parallel to these studies of spinor condensates, the efforts are multi-component
quantum gases have also received equal attention, most notably the case of
pseudo spin-1/2 condensates \cite{hall98,thalhammer08,papp08,kuklov02,ashhab03,shi06}
involving two atomic species or components. Since the total spin is conserved
under the density dependent interaction between the two components, even in the
presence of evaporative cooling if the interaction is SU(2) symmetric,
A. B. Kuklov {\it et al.} \cite{kuklov02} predicted that in the ground states
all atoms condense into two orthogonal spatial orbitals, resulting in a
maximally entangled many atom state. Replacing the two orbitals
with two different species of atoms, Shi {\it et al.} \cite{shi06} found
a ground state condensate with an entangled order parameter.
For two bosonic species of spin-1 atoms, such as $^{87}$Rb and $^{23}$Na atoms
in the lowest hyperfine state manifold, the mixture reduces to a binary mixture of spin-1 BECs.
In the absence of external B-field, the ground state phases are essentially known,
from the semi-classical phase diagram under the mean-field approximation \cite{xu09}
to the quantum results under the SMA for each of the two species
and including quantum spin fluctuations \cite{xu10,zhang10,shi09}.
Several interesting results have been revealed already.
For example, the ground state reduces to a maximally entangled states \cite{xu10,shi09}
where the mixture becomes fragmented and the number fluctuations \cite{zhang10} exhibit drastically
different features from the isolated spin-1 condensates, when the inter-species'
anti-ferromagnetic spin exchange interaction is large enough and when
the inter-species singlet pairing interaction is ignored.

In this work, we continue our study of a binary mixture of spin-1 condensates,
in the presence of a weak external B-field. First, as in all experimental setups,
a nonzero B-field inevitably exists no matter how small it may be. Second,
it is motivated by the possibilities of interesting physics due to the
competitions between the linear and quadratic Zeeman shifts and the ferromagnetic
interactions, which for the case of isolated spin-1 condensate
is shown to induce a spontaneous axisymmetry breaking phase \cite{stenger98,murata07}.
We find that by tuning the inter-species spin exchange interaction and the linear
Zeeman shift, and largely ignoring the quadratic Zeeman shifts (because the B-field
is weak), there exists an analogous BA phase in a system of binary mixture of spin-1 condensates.

This article is organized as follows.
In section II, we introduce the model formulation for a binary mixture of
spin-1 condensates in the presence of an external B-field.
Sec. III is devoted to the mean field results of the broken-axisymmetry (BA) phase,
where we discuss the special case and the general case respectively.
In Sec. IV, we elaborate on the Bogoliubov spectra for the various
quantum phases of the mixture, and we point out the interesting zero
energy Goldstone modes responsible for restoring the broken continuous symmetries:
U(1) and SO(2). Also we propose a new scheme for classifying the
various phases in the ground states using the number of nonzero components
for their associated order parameters. Finally we conclude in Sec. V.
The appendix provides some mathematical details.

\section{Our model}

In this section, we first review the basic formulation
introduced earlier for the inter-species atomic interactions.
Between two distinguishable spin-1 atoms
(of different species), they
can be parameterized by the three different scattering lengths
${a}_{0,1,2}^{(12)}$ which refer to three channels of total
spin $F_{\rm tot}=0,1$, and $2$ respectively.
Define ${g}_{0,1,2}^{(12)}=4\pi\hbar^2{a}_{0,1,2}^{(12)}/\mu$
with $\mu=M_1M_2/(M_1+M_2)$ the reduced mass and
$M_l$ the mass for each atomic species ($l=1,2$),
the corresponding pseudo-potential is given by
${V}_{12}(\mathbf{r}_1-\mathbf{r}_2)=({g}_0^{(12)}\mathcal{P}_0
+{g}_1^{(12)}\mathcal{P}_1+{g}_2^{(12)}\mathcal{P}_2)\delta(\mathbf{r}_1-\mathbf{r}_2)/2$.
 $\mathcal{P}_{0,1,2}$ is the corresponding
total spin projection operator, and
$\vec{F}_1\cdot\vec{F}_2=\mathcal{P}_2-\mathcal{P}_1-2\mathcal{P}_0$.
The inter-species interaction is then expressed as
\begin{eqnarray}
  {V}_{12}(\mathbf{r}_1-\mathbf{r}_2)=\frac{1}{2}(\alpha+\beta\mathbf{F}_1\cdot\mathbf{F}_2+\gamma\mathcal{P}_0)
  \delta(\mathbf{r}_1-\mathbf{r}_2),
  \label{intmsc}
\end{eqnarray}
where $\alpha=({g}_1^{(12)}+{g}_2^{(12)})/2$,
$\beta=(-{g}_1^{(12)}+{g}_2^{(12)})/2$, and
$\gamma=(2{g}_0^{(12)}-3{g}_1^{(12)}+{g}_2^{(12)})/2$.
Denoting $\hat{\Psi}_{M_F}(\mathbf{r})$ and
$\hat{\Phi}_{M_F}(\mathbf{r})$ as the two species annihilation
field operator at a position $\mathbf{r}$ respectively,
the Hamiltonian of a homogenous system of binary mixture under B-field becomes
\begin{eqnarray}
  \hat{H}&=&\hat{H}_1+\hat{H}_2+\hat{H}_{12},\nonumber\\
  \hat{H}_1&=&\int d\mathbf{r}\,\left\{\hat{\Psi}^{\dag}_{m}
  \Big(-\frac{\hbar^2}{2M_1}\nabla^2-p_1 m+q_1 m^2 \Big)\hat{\Psi}_{m}\right. \nonumber\\
  &&\left.+\frac{\alpha_1}{2}
  \hat{\Psi}_i^{\dag}\hat{\Psi}_j^{\dag}\hat{\Psi}_j\hat{\Psi}_i
  +\frac{\beta_1}{2}\hat{\Psi}_i^{\dag}\hat{\Psi}_k^{\dag}
  \mathbf{F}_{1ij}\cdot\mathbf{F}_{1kl}\hat{\Psi}_l\hat{\Psi}_j\right\},\nonumber\\
  \hat{H}_{12}&=&\frac{1}{2}\int d\mathbf{r}\,\left\{\alpha
  \hat{\Psi}_i^{\dag}\hat{\Phi}_j^{\dag}\hat{\Phi}_j\hat{\Psi}_i
  \right.\nonumber\\
  &&\left.+\beta\hat{\Psi}_i^{\dag}\hat{\Phi}_k^{\dag}\mathbf{F}_{1ij}\cdot
  \mathbf{F}_{2kl}\hat{\Phi}_l\hat{\Psi}_j
  +\frac{1}{3}\gamma\, \hat{s}^{\dag}\hat{s}\right\},
  \label{hamiltonian}
\end{eqnarray}
where $p_1$ and $q_1$ are the linear and the quadratic Zeeman shifts,
respectively, and summation over repeated indices are assumed.
$\hat{H}_2$ has the same expression as $\hat{H}_1$ except for the substitution
of subscript 1 by 2 and $\hat{\Psi}$ by $\hat{\Phi}$.
$\hat{s}=(\hat{\Psi}_1\hat{\Phi}_{-1}-\hat{\Psi}_0\hat{\Phi}_{0}
+\hat{\Psi}_{-1}\hat{\Phi}_{1})$ is the inter-species singlet pairing operator.

\section{Mean-field ground states}

This study concerns the limit of weak external B-fields, when the
quadratic Zeeman shifts can be neglected, or $q_1=q_2=0$.
Without loss of generality we assume $p_1p_2\ge0$.
For the mixture of two alkali species, such as the
spin-1 $^{87}$Rb and $^{23}$Na condensates, under the assumption of
interspecies interaction mainly arise from the contributions of the two
valence electrons \cite{weiss03,pashov05,stoof88}, we can approximate
$\gamma=0$ \cite{luo07}.
For a uniform system, it is convenient to expand the field operators
in terms of plane waves as $\hat{\Psi}_m=\Omega^{-1/2}\sum_\mathbf{k}
e^{i\mathbf{k}\cdot \mathbf{r}}\hat{a}_{\mathbf{k},m}$
and $\hat{\Phi}_m=\Omega^{-1/2}\sum_\mathbf{k}
e^{i\mathbf{k}\cdot \mathbf{r}}\hat{b}_{\mathbf{k},m}$,
where $\Omega$ is the system volume, and $\hat{a}_{\mathbf{k},m}$
($\hat{b}_{\mathbf{k},m}$) denotes the annihilation operator of
an atom of species one (two) with momentum $\hbar\mathbf{k}$ and magnetic
quantum number $m$. Then the Hamiltonian of Eq. (\ref{hamiltonian})
can be rewritten as
\begin{eqnarray}
  \hat{H}_1&=&\sum\limits_{\mathbf{k},m}(\varepsilon_{1\mathbf{k}}-p_1m)
  \hat{a}^{\dag}_{\mathbf{k},m}\hat{a}_{\mathbf{k},m}
  +\frac{\alpha_1}{2\Omega}\sum\limits_{\mathbf{k}}:\hat{\rho}_{1,\mathbf{k}}
  \hat{\rho}_{1,-\mathbf{k}}:\nonumber\\
  &&+\frac{\beta_1}{2\Omega}\sum\limits_{\mathbf{k}}:\hat{\mathbf{f}}_{1,\mathbf{k}}
  \cdot \hat{\mathbf{f}}_{1,-\mathbf{k}}:,\nonumber\\
  \hat{H}_{12}&=&\frac{\alpha}{2\Omega}\sum\limits_{\mathbf{k}}\hat{\rho}_{1,\mathbf{k}}
  \hat{\rho}_{2,-\mathbf{k}}+\frac{\beta}{2\Omega}\sum\limits_{\mathbf{k}}
  \hat{\mathbf{f}}_{1,\mathbf{k}}\cdot\hat{\mathbf{f}}_{2,-\mathbf{k}},
  \label{hamiltonian2}
\end{eqnarray}
where $\varepsilon_{1\mathbf{k}}=\hbar^2\mathbf{k}^2/2M_1$, $\hat{\rho}_{1,\mathbf{k}}
=\sum_{\mathbf{q},m}\hat{a}^{\dag}_{\mathbf{q}+\mathbf{k},m}\hat{a}_{\mathbf{q},m}$,
and $\hat{\mathbf{f}}_{1,\mathbf{k}}
=\sum_{\mathbf{q},m,n}\hat{a}^{\dag}_{\mathbf{q}+\mathbf{k},m}\mathbf{F}_{1mn}\hat{a}_{\mathbf{q},n}$.
The symbol $:\ :$ represents the normal ordering of the operators.
Similar to Eq. (\ref{hamiltonian}), $\hat{H}_2$ is given by
replacing the subscript 1 by 2, and the operator $\hat{a}$ by $\hat{b}$ in $\hat{H}_1$.

The condensate component corresponds to the zero
momentum $\mathbf{k}=0$ state, which is occupied by
a macroscopic number of atoms. The corresponding Hamiltonian becomes
\begin{eqnarray}
  \hat{H}_{\rm BEC}&=&\sum\limits_m -m (p_1\hat{a}_{\mathbf{0},m}^{\dag}
  \hat{a}_{\mathbf{0},m}+p_2\hat{b}_{\mathbf{0},m}^{\dag}
  \hat{b}_{\mathbf{0},m})\nonumber\\
  &+&\frac{\alpha_1}{2\Omega}:\hat{\rho}_{1,\mathbf{0}}^2:
  +\frac{\beta_1}{2\Omega}:\hat{\mathbf{f}}_{1,\mathbf{0}}^2:
  +\frac{\alpha_2}{2\Omega}:\hat{\rho}_{2,\mathbf{0}}^2:
  \nonumber\\
  &+&\frac{\beta_2}{2\Omega}:\hat{\mathbf{f}}_{2,\mathbf{0}}^2:
  +\frac{\alpha}{2\Omega}\hat{\rho}_{1,\mathbf{0}}\hat{\rho}_{2,\mathbf{0}}
  +\frac{\beta}{2\Omega}\hat{\mathbf{f}}_{1,\mathbf{0}}\cdot\hat{\mathbf{f}}
  _{2,\mathbf{0}}. \hskip 24pt
  \label{hambec}
\end{eqnarray}
Under the mean-field approximation,
the operator $\hat{a}_{\mathbf{0},m}$ ($\hat{b}_{\mathbf{0},m}$)
is replaced by $c$-number $\sqrt{N}_1\zeta_{1,m}$ ($\sqrt{N}_2
\zeta_{2,m}$), where $N_1$, $N_2$ are the atom numbers for the species one
and two respectively, and $\zeta_{1}=(\zeta_{1,1},\zeta_{1,0},\zeta_{1,-1})^T$
and $\zeta_{2}=(\zeta_{2,1},\zeta_{2,0},\zeta_{2,-1})^T$ are normalized spin-1 spinors.
As a result, in the mean-field approximation the ground state spinor wavefunction
is found by minimizing the mean-field spin-dependent energy given by
\begin{eqnarray}
  \mathcal{E}_s&=&
  -p_1 \langle F_{\rm 1z}\rangle+\frac{1}{2}\beta_1 n_1 \langle \mathbf{F}_{1}\rangle^2
  -p_2 \frac{N_2}{N_1}\langle F_{\rm 2z}\rangle
  \nonumber\\
  &&+\frac{1}{2}\beta_2 n_2\frac{N_2}{N_1}\langle \mathbf{F}_2\rangle^2
  +\frac{1}{2}\beta n_2\langle \mathbf{F}_1\rangle
  \cdot\langle\mathbf{F}_2\rangle\nonumber\\
  &=&-pf_{\rm 1z}-xpf_{\rm 2z}+\frac{1}{2}(\beta'_1\mathbf{f}_1^2
  +\beta'_2\mathbf{f}_2^2+\beta'\mathbf{f}_1\cdot\mathbf{f}_2), \hskip 24pt
  \label{sdenergy}
\end{eqnarray}
where
$n_j=N_j/\Omega$ is the condensate density
of the $j$-th species and
$\mathbf{f}_{1,2}=\langle \mathbf{F}_{1,2}\rangle$,
$p_1=p_2N_2/(N_1x)=p$, $\beta'_1=\beta_1 n_1$,
$\beta'_2=\beta_2n_2 N_2/N_1$ and $\beta'=\beta n_2$.
The procedure to find the mean-field ground states is
summarized in the appendix \ref{appendix},
where we discuss the structure of the ground
states classified into two classes: with or without broken-axisymmetry.
According to Ref. \cite{ueda10}, a state with a nonzero transverse magnetization
is called a broken-axisymmetry phase. Analogously states preserving
axisymmetry refer to those with zero transverse magnetization.
In the BA phase, both species are found to be fully polarized.
For axisymmetry preserved
ground states, the wave functions take the form
\begin{eqnarray}
  \zeta_j=
  e^{i\chi_{j}}\left(
  \begin{array}{c}
    e^{i\varphi_{j}}\sqrt{(1+f_{\rm jz})/2} \\
    0 \\
    \sqrt{(1-f_{\rm jz})/2} \\
  \end{array}
  \right),
  \label{opaxispreserved}
\end{eqnarray}
where $\chi_j$ and $\varphi_j$ are arbitrary phase angles, and $j=1,2$.
When $f_{\rm 1z}=0$ or $f_{\rm 2z}=0$, the ground states
include an infinite family of degenerate states with that in Eq.~(\ref{opaxispreserved}).
More rigorously, we note that the nematic order
for the above state (\ref{opaxispreserved}) actually is not
axisymmetric \cite{ueda10}.
For the BA phase, the ground state wave functions take the form
\begin{eqnarray}
  \zeta_j=
  e^{i\chi_{j}}\left(
  \begin{array}{c}
    e^{-i\varphi_j}\cos^2\frac{\theta_j}{2} \\
    \sqrt{2}\cos\frac{\theta_j}{2}\sin\frac{\theta_j}{2} \\
    e^{i\varphi_j}\sin^2\frac{\theta_j}{2} \\
  \end{array}
  \right),
  \label{opaxisbroken}
\end{eqnarray}
where $\chi_j$ and $\varphi_j$ are arbitrary phase angles, and $\varphi_2=\varphi_1+\pi (\text{mod}\ 2\pi)$,
$j=1,2$. $\theta_1$ and $\theta_2$ are determined by $f_{\rm 1z}$ and $f_{\rm 2z}$
respectively as $\cos\theta_1=f_{\rm 1z}$ and $\cos\theta_2=f_{\rm 2z}$.
The ground state phase diagram is
characterized into three cases according to the intra-species
spin exchange interaction parameter $\beta'_1$ and $\beta'_2$.
Without loss of generality, in the following we assume $|\beta'_1|<|\beta'_2|$.

\subsection{the special case of $x=1$}

\begin{figure}[tbp]
\centering
\includegraphics[width=3.0in]{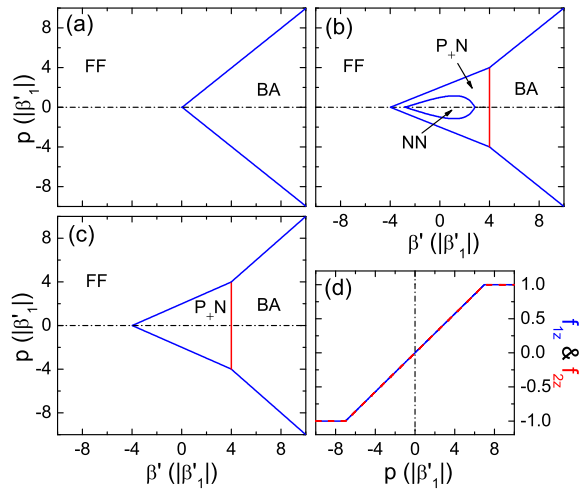}
\caption{(Color online). The ground state phase diagram of
our model system of a binary spin-1 condensate mixture
at fixed values of $\beta'_1$ and $\beta'_2$ for the special
case of $x=1$.
The black dash-dotted lines correspond to $p=0$,
which serve as guides for the eye. Blue (Red) lines
denote continuous (discontinuous) phase boundaries.
The first three subplots denote fixed intra-specie spin exchange
interaction parameters at
($\beta'_1$, $\beta'_2$)/$|\beta'_1|=$: (a) ($-1$,$-2$); (b)
($1,2$); and (c) ($-1,2$). The last one (d) illustrates the
dependence of ground state order parameters $f_{\rm 1z}$ (blue solid line)
and $f_{\rm 2z}$ (red dashed line)
of the BA phase on the linear Zeeman shift $p$ at a fixed value of
$\beta'=7|\beta'_1|$. }
\label{fig1}
\end{figure}

In this subsection, we consider the special case of $x=1$,
where the parameter defined as the partical number multiplying the linear Zeeman shift,
of the two species are equal ($N_1p_1=N_2p_2$).
The ground states are as shown in the Fig.~(\ref{fig1}),
where the first three subplots denote
fixed intra-specie spin exchange interaction
parameters at ($\beta'_1$, $\beta'_2$)/$|\beta'_1|=$: (a) ($-1$,$-2$); (b)
($1,2$); and (c) ($-1,2$). The last one (d) illustrates the
dependence of ground state order parameters $f_{\rm 1z}$ and $f_{\rm 2z}$
of the BA phase on the linear Zeeman shift $p$ at a fixed value of
$\beta'=7|\beta'_1|$.

\begin{figure}[tbp]
\centering
\includegraphics[width=2.2in]{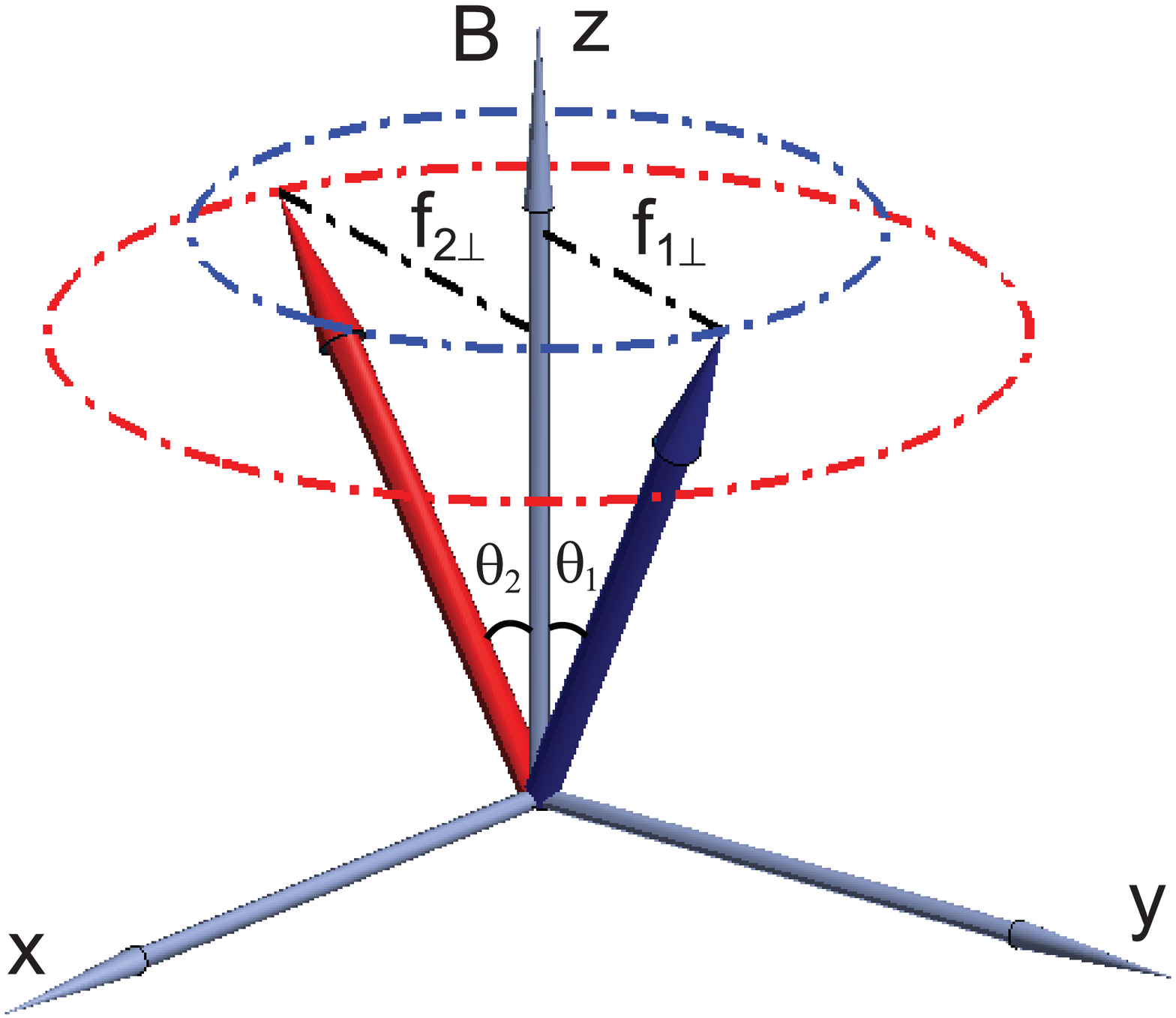}
\caption{(Color online). A schematic illustration of the spin
vectors for both condensate species in the broken-axisymmetry phase.
The weak external B-field is along the $z$-axis direction.
The blue (red) arrow denotes the spin vector of species one (two),
which tilts at an angle with respect to the $z$-axis.
The two spin vectors
and the $z$-axis are in the same plane. The blue and red dotted dash lines
correspond to the orbits of the end points of two spin vectors in the degenerate
ground state of the broken-axisymmetry phase. The black dotted dash lines
represent the transverse magnetizations $f_{1\perp}$ and $f_{2\perp}$ of
two species respectively.}
\label{fig2}
\end{figure}

First, we discuss the case when the two spin-1 condensates
are both ferromagnetic ($\beta'_1<0$ and $\beta'_2<0$).
The ground state has two phases: the FF phase and the BA phase.
The FF phase is a typical axisymmetry preserving phase
with $f_{\rm 1z}=f_{\rm 2z}=\text{sign}(p)$, which is
the same as that found in \cite{xu09} with atomic spins in each species
fully polarized and aligned parallel to each other with
$\mathbf{f}_1^2=\mathbf{f}_2^2=\mathbf{f}_1\cdot\mathbf{f}_2=1$.
In the BA phase, the two spin vectors of each species are remain fully
polarized but now tilted at an angle $\theta_1$ and $\theta_2$
with $\theta_1=\theta_2=\arccos(p/\beta')$ with respect to the $z$-axis.
This is illustrated in the Fig.~\ref{fig2}.
From Eq.~(\ref{appderiv}), we confirm that in this case
the BA phase exists only when $\beta'\ge0$, which is consistent
with numerical and analytical results shown in the
Fig.~\ref{fig1}(a). The boundary between the FF phase
and the BA phase can be derived by comparing the energy of
Eq. (\ref{sdenergy}), which turns out to be $|p|=\beta'$.
In Fig.~\ref{fig1}(d), we present the dependence of the order parameters
$f_{\rm 1z}$ and $f_{\rm 2z}$ on the linear Zeeman shift $p$
at a fixed value of $\beta'=7|\beta'_1|$. In the region of
$|p|\le\beta'$, $f_{\rm 1z}$ and $f_{\rm 2z}$ change
smoothly from $-1$ to $1$ simultaneously.

We then turn to another case when one of the spin-1 condensates
is ferromagnetic ($\beta_1'<0$) and the other is antiferromagnetic or polar ($\beta'_2>0$).
In the absence of B-field, the competition between the intra- and inter-species spin
exchange interactions gives rise to three phases \cite{xu09}.  When
$\beta<-2\beta_2$, the ground state is in the FF phase.
Increasing the value of $\beta$, there arises another phase,
where the spin vector in the ferromagnetic one is fully polarized,
and the other in the polar one is not fully polarized, giving rise to
$\mathbf{f}_1^2=1$, $\mathbf{f}_2^2=(\beta'/2\beta'_2)^2$,
and $\mathbf{f}_1\cdot\mathbf{f}_2=-\beta/2\beta_2$.
In the other limit when $\beta'>2\beta'_2$,
the ground state is in the AA phase where the two spin vectors are fully polarized
into opposite directions with $\mathbf{f}_1^2=\mathbf{f}_2^2=1$
and $\mathbf{f}_1\cdot\mathbf{f}_2=-1$.
In the presence of a weak external B-field, these three phases for the ground state
in the phase space will extend into three regions.
The first one is the FF phase, which has the same properties
as that in the $p=0$ case, except now all spins are polarized along
the external B-field direction. The second one corresponds to the other axisymmetry
preserved phase denoted by $\rm P_+N$,
where we use symbol P and N to represent respectively the fully
and non-fully polarized spin vectors of two species.
The subscript $+$ is used to clarify that the fully
polarized spin is along the same direction of the B-field.
The other remaining one is the BA phase, which is totally the same
as that in the case of two ferromagnetic spin-1 condensates.
In Fig. \ref{fig1}(c), we illustrate the ground state
phase diagram. The boundary between the FF phase and the
BA phase is the same as that in Fig. \ref{fig1}(a),
except that when $\beta'<2\beta'_2$, the ground
state is changed to the $\rm P_+N$ phase, which also means
that the boundary between the $\rm P_+N$ phase and the BA phase
is $\beta'=2\beta'_2$, which is consistent with Eq.~(\ref{appderiv}).
Only when $\beta'\ge \max(0,2\beta'_1x,2\beta'_2/x)=2\beta'_2$
is satisfied will the BA phase possibly exist.  The remaining boundary
between the FF phase and the $\rm P_+N$ phase satisfies the
relationship $|p|=\beta'_2+\beta'/2$.
In the $\rm P_+N$ phase, we have $f_{\rm 1z}=\text{\rm sign}(p)$ and
$f_{\rm 2z}=(p-f_{\rm 1z}\beta'/2)/\beta'_2$.

Next, we consider the last case with two spin-1 polar condensates
($\beta'_2>\beta'_1>0$).
In Fig. \ref{fig1}(b), we present the phase diagram as well as
the boundaries separating the phases. Similar to the second case of
Fig. \ref{fig1}(c), the ground state has three phases: FF, $\rm P_+N$, and BA.
The order parameters and the boundaries between these three
phases are the same as that in the previous case of Fig. \ref{fig1}(c).
The only difference arises from the competition with the
intra-species interaction in the anti-ferromagnetic condensate.
A new phase emerges in the center of the $\rm P_+N$ phase,
which is denoted as the NN phase. In this new phase, neither spin vectors
of each species are fully polarized. The boundaries between the NN phase
and the $\rm P_+N$ phase satisfy the following relationships
$(\beta'/2-\beta'_2)|p|=\beta'^2/4-\beta'_1\beta'_2$.
For the ground states in the NN phase, we have
$f_{\rm 1z}=(\beta'/2-\beta'_2)p/(\beta'^2/4-\beta'_1\beta'_2)$ and
$f_{\rm 2z}=(\beta'/2-\beta'_1)p/(\beta'^2/4-\beta'_1\beta'_2)$.

\subsection{the general case}

\begin{figure}[tbp]
\centering
\includegraphics[width=3.0in]{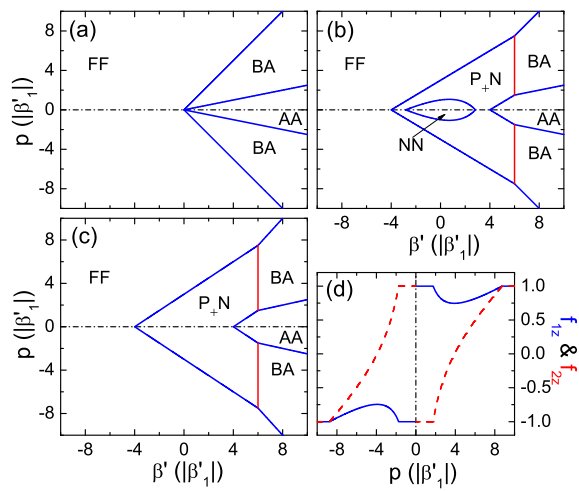}
\caption{(Color online). The same as in Fig.~\ref{fig1}, but
for $x=2/3$.}
\label{fig3}
\end{figure}

The ground state phase diagram for the special case of $x=1$ as discussed above
contains both axisymmetry preserved and broken phases. Their
boundaries can be partially determined by Eq.~(\ref{appderiv}).
The AA phase is found to exist only when $p=0$. It amounts to
an infinite set of degenerate states, where the
two spin vectors are antiparallel while remaining fully polarized.
As long as $x\ne1$ is satisfied, the region of the AA phase
extends to $p\ne0$, where one spin vector will
be parallel to the B-field, while the other one become antiparallel.
The order parameters ($f_{\rm 1z}, f_{\rm 2z}$) for the $\rm P_+N$, $\rm NN$ and BA phases
change into the following forms
$\big(\text{sign}(p), (2xp-\text{sign}(p)\beta')/2\beta'_2\big)$,
$\big( (2xp\beta'-4p\beta'_2)/(\beta'^2-4\beta'_1\beta'_2),(2p\beta'-4xp\beta'_1)/(\beta'^2-4\beta'_1\beta'_2)\big)$,
$\big(xp/\beta'-(x^2-1)\beta'/4xp, p/\beta'+(x^2-1)\beta'/4x^2p\big)$, respectively.

We first consider the case with $x<1$, whose ground state phase
diagram is demonstrated in Fig. \ref{fig3}. Due to our choice of
$|\beta'_1|<|\beta'_2|$, the ground state phase diagram is similar
to the case of $x=1$. One notable difference is that the AA phase appears
in the middle of the BA phase, and the boundaries between them
are given by $|p|=\beta'(1-x)/2x$. And the boundary between
the BA phase and FF phase is also changed to be described by $|p|=\beta'(x+1)/2x$.
As a result of the appearance of the AA phase, the $\rm P_+N$ phase
and the AA phase will now in touch with each other, and their boundary
is given by $|p|=(\beta'-2\beta'_2)/2x$.
The $\rm NN$ phase remains immersed in the $\rm P_+N$ phase,
and their boundary satisfies  $|p|=(\beta'^2-4\beta'_1\beta'_2)/(2x\beta'-4\beta'_2)$.
In Fig.~\ref{fig3}, we illustrate the phase diagram,
where the blue (red) solid lines denote the continuous (discontinuous)
phase boundaries. In Fig.~\ref{fig3}(d), we display the order
parameters of the ground states at fixed interspecies spin-exchange interaction
$\beta'=7|\beta'_1|$. Due to the emergence of the AA phase in the
middle of the BA phase, $f_{\rm 2z}$ changes from -1 to 1 twice
when $p$ changes from $-\infty$ to $\infty$, while $f_{\rm 1z}$
oscillates near the maximum and minimum values respectively,
which is different from the special case of $x$ discussed previously, where
$f_{\rm 1z}$ and $f_{\rm 2z}$ change simultaneously from -1 to 1.

\begin{figure}[tbp]
\centering
\includegraphics[width=3.0in]{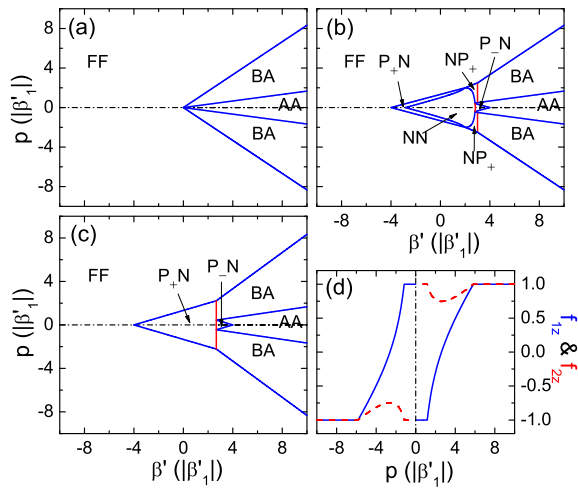}
\caption{(Color online). The same as in Fig.~\ref{fig1}, but
for $x=3/2$.}
\label{fig4}
\end{figure}

Next we discuss the other case of $x>1$, whose ground state phase
diagram is demonstrated in Fig. \ref{fig4}. Due to the choice of
$|\beta'_1|<|\beta'_2|$ and $x>1$, the ground state phase diagram is
much more complicated than that of $x\le1$. For two ferromagnetic
condensates ($\beta'_1<0$ and $\beta'_2<0$), the ground state phase
diagram is similar to that of $x<1$. With one ferromagnetic
($\beta'_1<0$) and one anti-ferromagnetic condensate ($\beta'_2>0$),
the phase diagram becomes different from the case of $x<1$. A new
phase denoted by $\rm P_-N$ emerges, where the subscript $-$ is used
to denote a spin vector being fully anti-aligned with the external
B-field.
This phase arises from increased interspecies spin-exchange
interaction $\beta'$ in this case of $x>1$. Starting from
the $\rm P_+N$ phase, with the increase of $\beta'$,
the spin vector of the second species becomes incresingly
polarized antiparallel to the B-field, which weights in more
linear Zeeman energy contribution from the second species,
and the spins for both species will flip to the opposite directions.
The order parameters for the $\rm P_-N$ phase
take the forms $f_{\rm 1z}=-\text{sign}(p)$ and $f_{\rm
2z}=(2xp+\text{sign}(p)\beta')/2\beta'_2$. Incidently, we find that
the phase boundaries between the $\rm P_+N$ phase and the BA or the
$\rm P_-N$ phases are the same: described by $\beta'=2\beta'_2/x$.
The remaining phase boundaries take the same forms as that in the
case of $x<1$, except for the one between the phase of $\rm P_-N$
and AA, which takes the form $|p|=(2\beta'_2-\beta')/2x$. For the
choice of two anti-ferromagnetic condensates ($\beta'_1>0$ and
$\beta'_2>0$), in addition to the $\rm P_-N$ phase, a new phase
denoted by $\rm NP_+$ emerges between the BA phase and the NN phase.
The order parameters for the $\rm NP_+$ phase are given by $f_{\rm
1z}=(2p-\text{sign}(p)\beta')/2\beta'_1$ and $f_{\rm
2z}=\text{sign}(p)$. The boundaries between the $\rm NP_+$ phase and
the NN, the AA, and the FF phases take the forma
$|p|=(\beta'^2-4\beta'_1\beta'_2)/(2\beta'-4x\beta'_1)$,
$|p|=\beta'/2-\beta'_1$, and $|p|=\beta'/2+\beta'_1$, respectively.
The complete phase diagram is illustrated in the Fig.~\ref{fig4}.
Additionally we illustrate the order parameters for the ground state
at fixed value of $\beta'=7|\beta'_1|$.  When $x>1$, $f_{\rm 1z}$ is
observed to change from -1 to 1 twice as $p$ changes from $-\infty$
to $\infty$, while $f_{\rm 2z}$ oscillates near the maximum and
minimum values, in contrast to the special case of $x<1$ as shown in
the Fig.~\ref{fig3}(d).

\section{Bogoliubov spectra and Goldstone modes}

This section is denoted to the discussion of Bogoliubov excitations of the
ground state phases we derived above. We follow the number-conserving
approach initially developed by Gardiner {\it et. al.} \cite{gardiner}
, where $\hat{a}_{\mathbf{0},m}$
($\hat{b}_{\mathbf{0},m}$) is replaced by $(N_1-\sum_{\mathbf{k}\ne0,m}\hat{a}^{\dag}
_{\mathbf{k},m}\hat{a}_{\mathbf{k},m})^{1/2}\zeta_{1,m}$
($(N_2-\sum_{\mathbf{k}\ne0,m}\hat{b}^{\dag}
_{\mathbf{k},m}\hat{b}_{\mathbf{k},m})^{1/2}\zeta_{2,m}$) \cite{castin,murata07,ueda00},
and keeping only quadratic terms of the operators to derive
the effective Hamiltonian by using the same notations as Refs. \cite{murata07,ueda00},
\begin{widetext}
\begin{eqnarray}
  \hat{H}^{\rm eff}&=&\hat{H}^{\rm eff}_1+\hat{H}^{\rm eff}_2+\hat{H}^{\rm eff}_{12},
  \nonumber\\
  \hat{H}^{\rm eff}_1&=&\sum\limits_{\mathbf{k}\ne0,m}\left(\varepsilon_{1\mathbf{k}}
  -p_1m+p_1\langle F_{\rm 1z}\rangle-\beta_1n_1\langle \mathbf{F}_1\rangle^2\right)\hat{a}_{\mathbf{k},m}^{\dag}
  \hat{a}_{\mathbf{k},m}+\beta_1n_1\langle\mathbf{F}_1\rangle\cdot\sum\limits_{\mathbf{k}\ne0,m,n}
  \hat{a}^{\dag}_{\mathbf{k},m}\mathbf{F}_{1mn}\hat{a}_{\mathbf{k},n}\nonumber\\
  &+&\frac{\alpha_1n_1}{2}\sum\limits_{\mathbf{k}\ne0}
  (2\hat{\mathcal{D}}_{1,\mathbf{k}}^{\dag}\hat{\mathcal{D}}_{1,\mathbf{k}}
  +\hat{\mathcal{D}}_{1,\mathbf{k}}\hat{\mathcal{D}}_{1,-\mathbf{k}}
  +\hat{\mathcal{D}}^{\dag}_{1,\mathbf{k}}\hat{\mathcal{D}}^{\dag}_{1,-\mathbf{k}})
  +\frac{\beta_1 n_1}{2}\sum\limits_{\mathbf{k}\ne0}(2\hat{\mathbf{\mathcal{F}}}^{\dag}_{1,\mathbf{k}}
  \cdot\hat{\mathbf{\mathcal{F}}}_{1,\mathbf{k}}+\hat{\mathbf{\mathcal{F}}}_{1,\mathbf{k}}\cdot
  \hat{\mathbf{\mathcal{F}}}_{1,\mathbf{-k}}+\hat{\mathbf{\mathcal{F}}}^{\dag}_{1,\mathbf{k}}\cdot
  \hat{\mathbf{\mathcal{F}}}_{1,\mathbf{-k}}^{\dag}),\nonumber\\
  \hat{H}^{\rm eff}_{12}&=&\frac{\beta}{2}\sum\limits_{\mathbf{k}\ne0,m}
  -\langle \mathbf{F}_1\rangle\cdot\langle \mathbf{F}_2\rangle
  (n_2\hat{a}^{\dag}_{\mathbf{k},m}\hat{a}_{\mathbf{k},m}+n_1\hat{b}^{\dag}_{\mathbf{k},m}\hat{b}_{\mathbf{k},m})
  +\frac{\beta n_1}{2}\langle\mathbf{F}_1\rangle\cdot\sum\limits_{\mathbf{k}\ne0,m,n}
  \hat{b}^{\dag}_{\mathbf{k},m}\mathbf{F}_{2mn}\hat{b}_{\mathbf{k,n}}
  \nonumber\\
  &+&\frac{\beta n_2}{2}
  \langle\mathbf{F}_2\rangle\cdot\sum\limits_{\mathbf{k}\ne0,m,n}
  \hat{a}^{\dag}_{\mathbf{k},m}\mathbf{F}_{1mn}\hat{a}_{\mathbf{k,n}}
  +\frac{\alpha \sqrt{n_1n_2}}{2}\sum\limits_{\mathbf{k}\ne0}
  (\hat{\mathcal{D}}^{\dag}_{1,\mathbf{k}}\hat{\mathcal{D}}_{2,\mathbf{k}}
  +\hat{\mathcal{D}}_{1,\mathbf{k}}\hat{\mathcal{D}}^{\dag}_{2,\mathbf{k}}
  +\hat{\mathcal{D}}_{1,\mathbf{k}}\hat{\mathcal{D}}_{2,-\mathbf{k}}
  +\hat{\mathcal{D}}^{\dag}_{1,\mathbf{k}}\hat{\mathcal{D}}^{\dag}_{2,-\mathbf{k}})
  \nonumber\\
  &+&\frac{\beta \sqrt{n_1n_2}}{2}\sum\limits_{\mathbf{k}\ne0}
  (\hat{\mathbf{\mathcal{F}}}^{\dag}_{1,\mathbf{k}}\cdot\hat{\mathbf{\mathcal{F}}}_{2,\mathbf{k}}
  +\hat{\mathbf{\mathcal{F}}}_{1,\mathbf{k}}\cdot\hat{\mathbf{\mathcal{F}}}^{\dag}_{2,\mathbf{k}}
  +\hat{\mathbf{\mathcal{F}}}_{1,\mathbf{k}}\cdot\hat{\mathbf{\mathcal{F}}}_{2,-\mathbf{k}}
  +\hat{\mathbf{\mathcal{F}}}^{\dag}_{1,\mathbf{k}}\cdot\hat{\mathbf{\mathcal{F}}}^{\dag}_{2,-\mathbf{k}})
  +\mathcal{E}_0,
  \label{hambogoliubov}
\end{eqnarray}
\end{widetext}
where $\mathcal{E}_0$ is a constant term and we have defined
$\hat{\mathcal{D}}_{1,\mathbf{k}}=\sum_{m}\zeta^*_{1,m}\hat{a}_{\mathbf{k},m}$,
$\hat{\mathbf{\mathcal{F}}}_{1,\mathbf{k}}=\sum_{m,n}\zeta^*_{1,m}\mathbf{F}_{1mn}\hat{a}_{\mathbf{k},n}$.
As before $\hat{H}^{\rm eff}_2$ takes the same form as $H^{\rm eff}_1$
except for the replacing of the subscript 1 by 2 and the operator $\hat{a}$ by $\hat{b}$.

Following the notation and approach of Ref. \cite{murata07},
we define operators
\begin{eqnarray}
  \hat{\mathbf{P}}_{\mathbf{k}}&=&(\hat{a}_{\mathbf{k},1},\hat{a}_{\mathbf{k},0},\hat{a}_{\mathbf{k},-1},
  \hat{b}_{\mathbf{k},1},\hat{b}_{\mathbf{k},0},\hat{b}_{\mathbf{k},-1})^T,\nonumber\\
  \hat{\mathbf{P}}^*_{\mathbf{k}}&=&(\hat{a}^{\dag}_{\mathbf{k},1},\hat{a}^{\dag}_{\mathbf{k},0},
  \hat{a}^{\dag}_{\mathbf{k},-1},\hat{b}^{\dag}_{\mathbf{k},1},\hat{b}^{\dag}_{\mathbf{k},0},
  \hat{b}^{\dag}_{\mathbf{k},-1})^T,
  \label{poperator}
\end{eqnarray}
which satisfy the Heisenberg equation
\begin{eqnarray}
  i\hbar\frac{d}{dt}\hat{\mathbf{P}}_{\mathbf{k}}=\mathcal{M}(k)\hat{\mathbf{P}}_{\mathbf{k}}
  +\mathcal{N}(k)\hat{\mathbf{P}}_{-\mathbf{k}}^*,
  \label{pheisenbergeq}
\end{eqnarray}
to calculate low-lying Bogoliubov modes.
$\mathcal{M}(k)$ and $\mathcal{N}(k)$ are chosen as real
with suitable real valued $\zeta_1$ and $\zeta_2$
and symmetric $6\times6$ matrix.
The operators for Bogoliubov quasi-particles are defined accordingly
\begin{eqnarray}
  \hat{\mathbf{Q}}_{\mathbf{k}}&=&
  (\hat{\mathcal{Q}}_{\mathbf{k},1},\hat{\mathcal{Q}}_{\mathbf{k},2},\hat{\mathcal{Q}}_{\mathbf{k},3},
  \hat{\mathcal{Q}}_{\mathbf{k},4},\hat{\mathcal{Q}}_{\mathbf{k},5},\hat{\mathcal{Q}}_{\mathbf{k},6})^T\nonumber\\
  &=&\mathcal{U}(k)\hat{\mathbf{P}}_{\mathbf{k}}+
  \mathcal{V}(k)\hat{\mathbf{P}}_{-\mathbf{k}}^*,
  \label{qoperator}
\end{eqnarray}
which enforces a diagonalized effective Hamiltonian
\begin{eqnarray}
  \hat{H}^{\rm eff}=\sum\limits_{\mathbf{k},m}E_{m}\hat{\mathcal{Q}}^{\dag}_{\mathbf{k},m}
  \hat{\mathcal{Q}}_{\mathbf{k},m}
  +E_{\rm vac},
  \label{hambogoliubov2}
\end{eqnarray}
where $E_{\rm vac}$ is the vacuum state energy for the quasi-particles.

As the quasi-particle operators $\hat{\mathcal{Q}}_{\mathbf{k},m}$ and
satisfy Bose commutation relations, we obtain
\begin{eqnarray}
  i\hbar\frac{d}{dt}\hat{\mathbf{Q}}_{\mathbf{k}}=E(\mathbf{k})\hat{\mathbf{Q}}_{\mathbf{k}},
  \label{qheisenbergeq}
\end{eqnarray}
where $E(\mathbf{k})$ is a diagonal $6\times6$ matrix: ${\rm diag}(E)=(E_{1},E_{2},E_{3},
E_{4},E_{5},E_{6})^T$.
The eigenvalues of the quasi-particles are found from solving
the following eigenequation \cite{You}
\begin{eqnarray}
  (\mathcal{M}+\mathcal{N})(\mathcal{M}-\mathcal{N})(\mathcal{U}+\mathcal{V})^T
  =(\mathcal{U}+\mathcal{V})^TE^2.
  \label{geigenequations}
\end{eqnarray}

We choose to reexamine the ground states of the Hamiltonian of Eq.~(\ref{hambec})
before discussing the Bogoliubov excitations and the Goldstone modes.
As shown in the appendix~\ref{appendix}, there exist two types of
mean field phases in the ground state, one preserves axisymmetry,
and the other breaks axisymmetry.
The axisymmetry preserved phases include FF, AA,
$\rm P_+N$, $\rm P_-N$, $\rm NP_+$, and NN phases, etc.
All ground state phases can be reclassified into four classes
according to the numbers of nonzero component in their order parameters:
the FF and AA phases falls into the two-component class,
the $\rm P_+N$, $\rm P_-N$, and $\rm NP_+$ phases belongs to the three-component class,
while the four-component class includes the NN phase and
the six component class includes the BA phase.

According to the Goldstone theorem: a gapless mode emerges
whenever a continuous symmetry is spontaneously broken.
For the two-component class, two U(1) symmetries for the two atomic species are broken \cite{Lyou}.
One and two more spin rotation symmetries are broken in both
the three-component and four-component classes. For the BA phase,
two U(1) symmetries and one SO(2) symmetry are broken \cite{Yi}.
To verify the above arguments, we solve the eigen-equations of Eq. (\ref{geigenequations})
employing both analytic or numerical techniques. In the ground state
of the Hamiltonian of Eq. (\ref{hambec}), unequal densities of two species
can modify the spin-exchange interaction and linear Zeeman shifts.
Without loss of generality, in the following numerical calculation,
we will assume $N_1=N_2$ and $n_1=n_2=n$. Because spin-independent
interactions do not change the general properties of the Bogoliubov excitations,
we will set intra- and inter-species spin-independent interaction as
$\alpha_1n_1=\alpha_2n_2=\alpha \sqrt{n_1n_2}=\alpha n=20 |\beta'_1|$
and we further assume $M_2=3M_1$.

\subsection{two-component class}

As summarized above, the two-component class contains two phases:
the FF phase and AA phase. We first consider the FF phase by assuming
$\zeta_1=\zeta_2=(1,0,0)^T$ and $p>0$. After detailed
calculation, we find
\begin{widetext}
\begin{eqnarray}
  E_{1}^2&=&\frac{1}{2}\Big(\epsilon_{1\mathbf{k}}^2+\epsilon_{2\mathbf{k}}^2
  -\sqrt{(\epsilon_{1\mathbf{k}}^2-\epsilon_{2\mathbf{k}}^2)^2+4(\alpha n+\beta n)^2
  \varepsilon_{1\mathbf{k}}\varepsilon_{2\mathbf{k}} }\Big),\nonumber\\
  E_{2}^2&=&\frac{1}{2}\Big(\epsilon_{1\mathbf{k}}^2+\epsilon_{2\mathbf{k}}^2
  +\sqrt{(\epsilon_{1\mathbf{k}}^2-\epsilon_{2\mathbf{k}}^2)^2+4(\alpha n+\beta n)^2
  \varepsilon_{1\mathbf{k}}\varepsilon_{2\mathbf{k}} }\Big),\nonumber\\
  E_{3}^2&=&\frac{1}{4}\Big(\varepsilon_{1\mathbf{k}}+\varepsilon_{2\mathbf{k}}-\beta n+p_1+p_2
  -\sqrt{(\varepsilon_{1\mathbf{k}}-\varepsilon_{2\mathbf{k}}+p_1-p_2)^2+(\beta n)^2}\Big)^2,\nonumber\\
  E_{4}^2&=&\frac{1}{4}\Big(\varepsilon_{1\mathbf{k}}+\varepsilon_{2\mathbf{k}}-\beta n+p_1+p_2
  +\sqrt{(\varepsilon_{1\mathbf{k}}-\varepsilon_{2\mathbf{k}}+p_1-p_2)^2+(\beta n)^2}\Big)^2,\nonumber\\
  E_{5}^2&=&(\varepsilon_{1\mathbf{k}}-2\beta_1n-\beta n+2p_1)^2,\nonumber\\
  E_{6}^2&=&(\varepsilon_{2\mathbf{k}}-2\beta_2n-\beta n+2p_2)^2,
  \label{eigsffphase}
\end{eqnarray}
\end{widetext}
where $\epsilon_{1\mathbf{k}}^2=\varepsilon_{1\mathbf{k}}^2+2(\alpha_1n+\beta_1n)\varepsilon_{1\mathbf{k}}$
and $\epsilon_{2\mathbf{k}}^2=\varepsilon_{2\mathbf{k}}^2+2(\alpha_2n+\beta_2n)\varepsilon_{2\mathbf{k}}$.
From Eq. (\ref{eigsffphase}), we confirm that there are two gapless
Goldstone modes with eigenvalues $E_1$ and $E_2$, associated
with the two coupled U(1) symmetry breaking modes between
the two $M_F=1$ components of two species respectively.
The two coupled modes between the two $M_F=0$ components are gapped, and
their excitation spectra are denoted by $E_{3}$ and $E_{4}$.
The remaining two excitation spectra are the two modes coupling between
two $M_F=-1$ components, which are gapped as well.
In the boundary between the FF phase and the BA phase, where $|p|=\beta n$,
we find the excitation spectra of $E_3$ reduce to a free
particle form at small values of momentum $k$.

For the AA phase, without loss of generality, we can
assume the order parameters of the forms
$\zeta_1=(1,0,0)^T$ and $\zeta_2=(0,0,1)^T$.
The excitation spectra can be analytically retrieved in this case as
\begin{widetext}
\begin{eqnarray}
  E_{1}^2&=&\frac{1}{2}\Big(\epsilon_{1\mathbf{k}}^2+\epsilon_{2\mathbf{k}}^2
  -\sqrt{(\epsilon_{1\mathbf{k}}^2-\epsilon_{2\mathbf{k}}^2)^2+4(\alpha n-\beta n)^2
  \varepsilon_{1\mathbf{k}}\varepsilon_{2\mathbf{k}} }\Big),\nonumber\\
  E_{2}^2&=&\frac{1}{2}\Big(\epsilon_{1\mathbf{k}}^2+\epsilon_{2\mathbf{k}}^2
  +\sqrt{(\epsilon_{1\mathbf{k}}^2-\epsilon_{2\mathbf{k}}^2)^2+4(\alpha n-\beta n)^2
  \varepsilon_{1\mathbf{k}}\varepsilon_{2\mathbf{k}} }\Big),\nonumber\\
  E_{3}^2&=&\frac{1}{4}\Big(\varepsilon_{1\mathbf{k}}-\varepsilon_{2\mathbf{k}}+p_1+p_2
  -\sqrt{(\varepsilon_{1\mathbf{k}}+\varepsilon_{2\mathbf{k}}+\beta n+p_1-p_2)^2-(\beta n)^2}\Big)^2,\nonumber\\
  E_{4}^2&=&\frac{1}{4}\Big(\varepsilon_{1\mathbf{k}}-\varepsilon_{2\mathbf{k}}+p_1+p_2
  +\sqrt{(\varepsilon_{1\mathbf{k}}+\varepsilon_{2\mathbf{k}}+\beta n+p_1-p_2)^2-(\beta n)^2}\Big)^2,\nonumber\\
  E_{5}^2&=&(\varepsilon_{1\mathbf{k}}-2\beta_1 n+\beta n+2p_1)^2,\nonumber\\
  E_{6}^2&=&(\varepsilon_{2\mathbf{k}}-2\beta_2 n+\beta n-2p_2)^2.
  \label{eigsaaphase}
\end{eqnarray}
\end{widetext}
Again there are two gapless Goldstone modes,
associated with the breaking of the two coupled U(1) symmetries
between the components of $\zeta_{1,1}$ and $\zeta_{2,-1}$ denoted by $E_1$ and $E_2$.
The two coupled gapped modes of the $M_F=0$ spin components for the two species
are denoted as $E_3$ and $E_4$. The remaining two gapped modes come from
the coupling between the components of $\zeta_{1,-1}$ and $\zeta_{2,1}$.
All results are consistent with our previous discussions.

\subsection{three-component class}

\begin{figure}[tbp]
\centering
\includegraphics[width=3.0in]{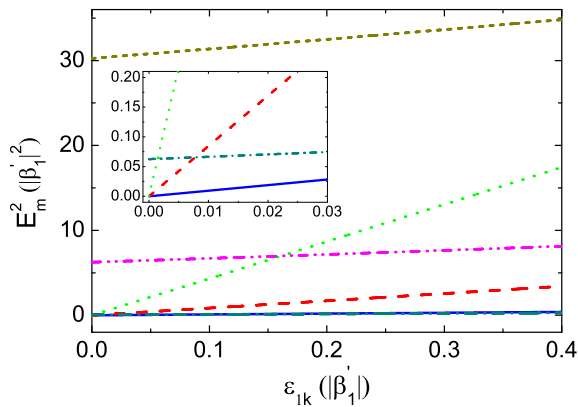}
\caption{(Color online). The Bogoliubov excitation spectra
for the $\rm P_+N$ phase.
The intra- and inter-species
spin-dependent interaction parameters used are
$\beta'_1<0$, $\beta'_2=2|\beta'_1|$, and $\beta'=3|\beta'_1|$.
The B-field parameters are $p_1=p_2=2.5|\beta'_1|$.
The inset shows the small momentum $k$ region.}
\label{fig5}
\end{figure}

For the three-component class, the spin vector of one species is fully
polarized while the other one is partially polarized, resulting in
the breaking of two U(1) symmetries and one SO(2) spin rotation symmetry.
For all phases: $\rm P_+N$, $\rm P_-N$, and $\rm NP_+$,
their Bogoliubov spectra contain three gapless Goldstone modes.
We verify the above results numerically by diagonalizing Eq. (\ref{geigenequations}).
The case of the $\rm P_+N$ is shown for illustrative purposes in Fig.~\ref{fig5},
where we assume spin-dependent parameters $\beta'_1<0$, $\beta'_2=2|\beta'_1|$,
$\beta'=3|\beta'_1|$, and linear Zeeman shifts $p_1=p_2=2.5|\beta'_1|$.
The order parameters take the forms of Eq. (\ref{opaxispreserved}),
where now we have $f_{\rm 1z}=1$ and we choose $\chi_1=\chi_2=\varphi_1=\varphi_2=0$.

\subsection{four-component class}

\begin{figure}[tbp]
\centering
\includegraphics[width=3.0in]{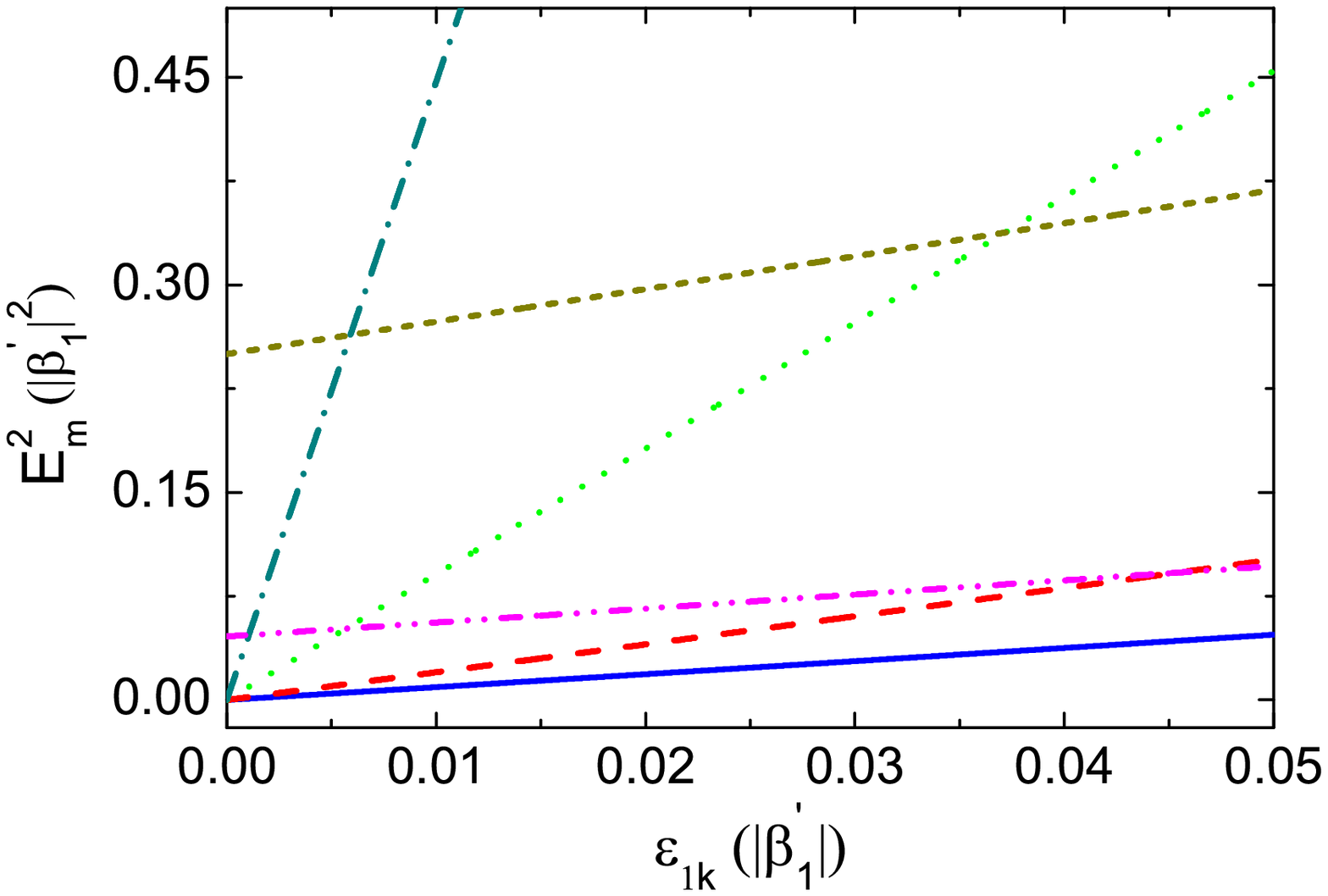}
\caption{(Color online). The same as in Fig. \ref{fig5}
but for the NN phase. The parameters adopted are
$\beta'_1>0$, $\beta'_2=2|\beta'_1|$, $\beta'=|\beta'_1|$,
and $p_1=p_2=|\beta'_1|/2$. }
\label{fig6}
\end{figure}

There exists only a single one phase: the NN phase, for the four-component class.
Its order parameter takes the form of Eq. (\ref{opaxispreserved}).
For simplicity, we assume $\chi_1=\chi_2=\varphi_1=\varphi_2=0$.
In Fig. \ref{fig6} we illustrates its collective excitation spectra,
where the intra- and inter-species spin-dependent interactions are
parameterized by $\beta'>0$, $\beta'_2=2|\beta'_1|$,
and $\beta'=|\beta'_1|$. The B-field parameters are chosen as $p_1=p_2=|\beta'_1|/2$.
Indeed one can infer that there are four gapless Goldstone modes in Fig. \ref{fig6},
consistent with the previous discussions.

\subsection{six-component class}

\begin{figure}[tbp]
\centering
\includegraphics[width=3.0in]{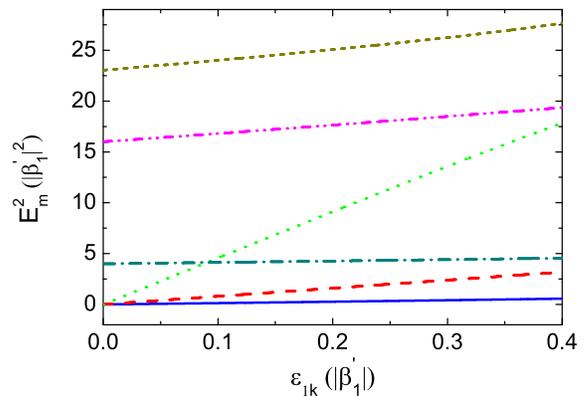}
\caption{(Color online). The same as in Fig. \ref{fig5}
but for the BA phase. The parameters adopted are
$\beta'_1>0$, $\beta'_2=2|\beta'_1|$, $\beta'=6|\beta'_1|$,
and $p_1=p_2=4\beta'/5$.}
\label{fig7}
\end{figure}

Finally for the six component class,
the axisymmetry is broken and its order parameter taking the form of Eq.~(\ref{opaxisbroken}).
Due to breaking of two U(1) and one SO(2) symmetries, the Bogoliubov excitation
spectra contain three gapless Goldstone modes, as confirmed by numerically
solving Eq.~(\ref{geigenequations}).
$\chi_1=\chi_2=\varphi_1=0$ and $\varphi_2=\pi$ are used for
numerical calculations, together with the
intra- and inter-species spin-dependent parameters
$\beta'_1>0$, $\beta'_2=2|\beta'_1|$, and $\beta'=6|\beta'_1|$.
The linear Zeeman shifts are assumed to satisfy $p_1=p_2=4\beta'/5$.
In Fig.~\ref{fig7}, we present all Bogoliubov excitation spectra,
from which one can easily verify the three gapless Goldstone modes.

\section{Conclusions}

In conclusion, we construct the ground state phase diagram
for a binary mixture of two spin-1 condensates under an
external weak B-field. Our results show that
the phase diagram can be classified into two categories: the axisymmetry
preserved and broken phases. The first one is shown to contain the FF, AA, $\rm P_+N$,
$\rm P_-N$, $\rm NP_+$, and NN phases, while the second one contains
only single phase: the BA phase, which appears only when
the inter-species anti-ferromagnetic spin-exchange interaction
is sufficiently large when compared to the intra-species
spin-exchange and linear Zeeman shifts.
Furthermore, according to the number of the gapless Goldstone modes,
the ground state phases can be classified into four classes:
two-, three-, four-, and six-component class, respectively accompanied by
two, three, four, and three gapless Goldstone modes.
We calculate all the Bogoliubov
excitation spectra with both analytical or numerical
techniques, and confirm our arguments given above.

Since the interspecies s-wave interaction parameters between two
spin-1 atoms, such as between $^{87}$Rb and $^{23}$Na atoms, are not
precisely known, we cannot conclude which of the above discussed
phases can be realized experimentally within presently available
experimental technologies. Given the promising innovation of tuning
interspecies interaction parameters with optical Feshbach resonances
and other tools, it seems that phases we study can stimulate
sufficient experimental interests to confirm their existence.

\section{Acknowledgments}

This work is supported by the NSF of China under Grant
Nos. 11004116, 10640420151 and 10974112 and by NKBRSF of China
under Grant Nos. 2006CB921206, 2006AA06Z104, 2006CB605105
and 2006CB921404.

\appendix
\section{constructing the ground state phases}
\label{appendix}

In this appendix, we give the detail discussion on how to clarify all possible ground state phases
for a binary mixture of spin-1 condensates under an
external weak B-field.

First, we assume that spontaneous axisymmetry breaking
doest not occur in either species, which results in $f_{1\pm}=f_{2\pm}=0$.
For species one, we then have
\begin{eqnarray}
  \sqrt{2}(\zeta^{(1)*}_1\zeta^{(1)}_0+\zeta^{(1)*}_0\zeta^{(1)}_{-1})=0,
  \label{f1peq0}
\end{eqnarray}
for the ground state. Without loss of generality, an unitary transformation
can be affected to set $\zeta^{(1)}_0$ as a real number. As a result,
there left only two solutions for Eq.~(\ref{f1peq0}):
$\zeta^{(1)}_0=0$ or $\zeta^{(1)*}_1+\zeta^{(1)}_{-1}=0$.
For the latter case, we infer that
$|\zeta^{(1)}_1|^2=|\zeta^{(1)}_{-1}|^2$ and $f_{\rm 1z}=0$,
which is a state degenerate with $\zeta_1=(e^{i\varphi_1},0,e^{i\varphi_{-1}})^T/\sqrt{2}$,
with arbitrary angles $\varphi_1$ and $\varphi_{-1}$.
Therefore the ground states is limited to the subspace with $\zeta^{(1)}_0=0$.
For the second species, an analogous reduction gives a similar subspace
with $\zeta^{(2)}_0=0$. Summarizing, without spontaneous axisymmetry
breaking, the ground states of the energy functional for a binary mixture
of two spin-1 condensates, Eq. (\ref{sdenergy}),
can be chosen as $\zeta_1=(\zeta_1^{(1)},0,\zeta_{-1}^{(1)})^T$
and $\zeta_2=(\zeta_1^{(2)},0,\zeta_{-1}^{(2)})^T$.

Next, we consider the structure of all possible spontaneous axisymmetry
breaking ground states. For a single spin-1 condensate,
the linear Zeeman shift does not induce BA phase.
If a spontaneous BA phase arises in a mixture,
both species must break axisymmetry. We therefore search
for the ground states of the BA phase
in the subspace of $\mathbf{f}_1^2=a^2$ and $\mathbf{f}_2^2=b^2$,
with $a^2\le1$ and $b^2\le1$. Since $p_1p_2\ge0$,
the BA phase appears for anti-ferromagnetic interspecies spin-exchange
interaction with $\beta'>0$.
In the ground states, the transverse magnetization of the two
species must align along opposite directions.

The order parameter for the BA phase
can be found from the following equations
\begin{eqnarray}
  \frac{d\mathcal{E}_s}{d f_{\rm 1z}}&=&-p+\frac{1}{2}\beta'\left(f_{\rm 2z}
  +\frac{\sqrt{b^2-f_{\rm 2z}^2}}{\sqrt{a^2-f_{\rm 1z}^2}}f_{\rm 1z}\right)=0,\nonumber\\
  \frac{d\mathcal{E}_s}{d f_{\rm 2z}}&=&-xp+\frac{1}{2}\beta'\left(f_{\rm 1z}
  +\frac{\sqrt{a^2-f_{\rm 1z}^2}}{\sqrt{b^2-f_{\rm 2z}^2}}f_{\rm 2z}\right)=0. \hskip 24pt
  \label{appba}
\end{eqnarray}
The solution of the above equations for the BA phase is given by
$f_{\rm 1z}=xp/\beta'+a^2\beta'/4xp-xb^2\beta'/4p$ and
$f_{\rm 2z}=p/\beta'+b^2\beta'/4p-a^2\beta'/4x^2p$.
We can calculate the first derivatives of the spin-dependent
energy function $\mathcal{E}_s$ with respect to $a^2$ and $b^2$
at the BA phase, and the results are
\begin{eqnarray}
  \frac{d \mathcal{E}_s}{da^2}&=&\frac{1}{2}\beta'_1-\frac{1}{2}\beta'
  \frac{\sqrt{b^2-f_{\rm 2z}^2}}{2\sqrt{a^2-f_{\rm 1z}^2}}
  =\frac{1}{4}(2\beta'_1-\beta'/x),\nonumber\\
  \frac{d \mathcal{E}_s}{db^2}&=&\frac{1}{2}\beta'_2-\frac{1}{2}\beta'
  \frac{\sqrt{a^2-f_{\rm 1z}^2}}{2\sqrt{b^2-f_{\rm 2z}^2}}
  =\frac{1}{4}(2\beta'_2-\beta'x). \hskip 24pt
  \label{appderiv}
\end{eqnarray}
We therefore conclude that the BA phase is possiblely exist
only when $\beta'\ge \max(0, 2\beta'_1 x, 2\beta'_2/x)$.
Because the first derivatives are both independent of $a^2$ and $b^2$,
the ground states of the BA phase will have $a^2=b^2=1$.

The actual ground state is found through a comparison of the
ground state energies for two types of phases: with/without broken-axisymmetry.
In addition, the results found here are subsequently affirmed with the numerical ones
from the use of simulated annealing method \cite{xu09} to minimize
the spin-dependent energy functional.

\end{document}